# Granular packing as model glass formers

**Yujie Wang(王宇杰)**[1,2,†]

[1] *Department of Physics and Astronomy, Shanghai Jiao Tong University,*
*800 Dong Chuan Road, Shanghai 200240, China*

[2] *Materials Genome Initiative Center, Shanghai Jiao Tong University,*
*800 Dong Chuan Road, Shanghai 200240, China*

Static granular packings are model hard-sphere glass formers. The nature of glass transition has remained a hotly debated issue. We review recent experimental progresses in using granular materials to study glass transitions. We focus on the growth of glass order with five-fold symmetry in granular packings and relate the findings to both geometric frustration and random first-order phase transition theories.

**Keywords: Granular systems, Disordered solids, Glasses, X-ray microscopy**

**PACS:** 45.70.-n, 61.43.-j, 61.43.Fs, 68.37.Yz

## 1. Introduction

The nature of dynamic arrest in glassy systems has remained a mystery and hotly debated despite years of research. Over the years, many theoretical approaches have been developed to understand its origin. The use of granular packings as model glass formers to study glass transition has a long history. Back in 1950s, Bernal has already used hard-sphere granular packing to simulate liquid and glass structures[1]. Until today, it remains a paradigm for metallic glass packing structures. This analogy was further explored by Edwards to relate glassy inherent states with mechanical stable granular packings[2]. Using the statistical framework based on Edwards' ensemble, concepts like effective temperature have greatly advanced our understanding of both granular physics and aging behavior in glass systems[3, 4]. Also, granular model system was originally believed to be a good model system for the study of avalanche and self-organized criticality (SOC)[5]. Other than being an important statistical model system to understand out-of-equilibrium and glassy behaviors, the understanding of granular materials will have far-reaching impacts on real-life applications as well since granular materials are omnipresent in industrial and geological processes like transportation of materials, landslide, and earthquake[6]. Therefore, a clear understanding of how granular materials can turn into amorphous solids and acquire mechanical rigidity will greatly help us understanding many of these processes. In this short review, we will review some of the recent researches on glassy behaviors in granular materials, mainly with a structural perspective. For a latest understanding of the progress on glass transition, there exist many excellent reviews [7-11]. Some of the major concepts and theories developed include energy landscape approach[12], free-volume theory[13], mode-coupling theory (MCT)[14], Adam-Gibbs theory[15], random first-order phase transition (RFOT)[16], jamming phenomenon[17], dynamic facilitation theory[18], and frustration limited domain theory[19], etc.

[†] Corresponding author. E-mail: yujiewang@sjtu.edu.cn

## 2. Structural mechanism of glass transition.
### 2.1. Structural order based on icosahedron

Geometric frustration has been long proposed to be one of the leading mechanisms for glass transition. Especially, local structures with five-fold symmetry like icosahedra have the best local packing but cannot tile space. The existence and importance of five-fold symmetric structures in liquids and glasses have been noted long time ago[20-22]. Later on, the geometric frustration concept was further developed into the frustration-limited domain theory[19] in which a criticality to icosahedral order in curved space is avoided when mapped back to Euclidean space. However, there also exists the possibility that glass transition could be related to frustrated crystalline or quasicrystalline orders as suggested by Tanaka[23] and the packing model by Miracle[24].

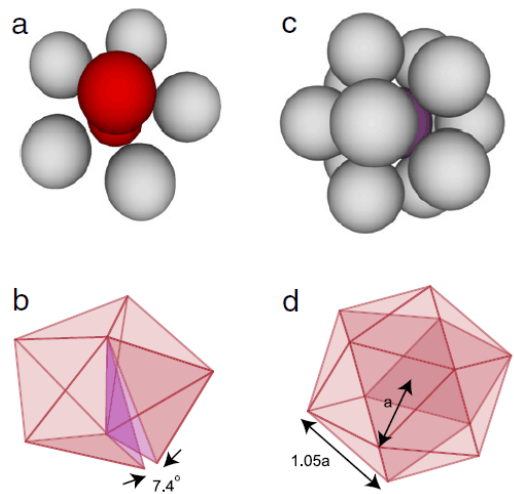

**Fig.1.** Frustration in polytetrahedra in Euclidean d = 3. (a) The pentagonal bipyramid ''7A'' is constructed with five tetrahedra (b) which leads to a gap of 7.4°. (c) 13-membered icosahedron. (d) Icosahedra have bonds between particles in the shell stretched by 5% with respect to the bonds between the central particle and the shell particles. The figure is reproduced from[25].

Experiments have been carried out to verify the existences of local structures with five-fold symmetry in liquids and glasses[26-28]. However, due to intrinsic difficulty, the relationship of icosahedron to slow dynamics can be carried out by simulation studies only. Efforts have been spent both on model and metallic glass formers in investigating the percolation of icosahedral structures[29-31], correlation between slow dynamics and icosahedral structures[32-35], soft spots and icosahedral structures[36, 37], *etc*. Additionally, local packing structures with five-fold symmetry, being only part of a full icosahedron, have also been found to play important roles in various amorphous systems[23, [38-41]. Since the icosahedron-like structures in glassy systems are never ideal, there currently exist many different methods to characterize them[31, 38, 42]. A detailed comparison of the similarities and differences among all these methods may help bridge different communities. Additionally, as noted in the following, most of the existing methods tend to ignore the specific ways how neighboring five-fold structures are connected.

**2.2. Structural order based on other definitions**

Despite the alluded relevance to the underlying structure by many theoretical approaches above, a direct "amorphous order" approach to glass transition based on specific structural motifs is not generally accepted [25]. Interestingly, this is to be compared with researchers working on applied glass formers like metallic glasses, who seem to accept a structural mechanism without too much hurdle[43]. The difficulty lies in the fact that it seems rather unlikely there exists a universal way to categorize different "orders" for so many diversified amorphous systems as we have done with crystalline materials, *i.e.,* we have to come up an icosahedron-like equivalent for different systems[20]. To avoid this search of an "order" based on physical intuition, many "order-agnostic" approaches have been developed which are based on information theory[44], graph theory[45] or recently developed machine-learning tools[46]. Especially, the pinning or "point-to-set" approach is originally motivated by the RFOT mosaic picture. It assumes that when the lengthscale of the pinning boundary or average distance between pinning sites is close to the size of the mosaic tile, the system can no longer relax[47]. However, as all above techniques statistically average many structural details, it is still doubtful how much they can outperform methods based on two-body correlation, local free volume or local entropy[48], which have been considered as sometimes useful but too crude in capturing the structural change upon dynamic arrest. Structural analyses based on high-order structural correlation functions like bond-orientational order[49], Voronoi polyhedra[43], common neighbor analysis[50], or local favored structures[51] are considered to be more useful in this sense despite their less flexibility.

**2.3. Structural and dynamical order**

The dynamic slowdown of a glass transition is accompanied not only by a week growth of static correlation length, but also a dynamic one [52]. Kob *et al.* found a non-monotonic temperature of dynamic correlation length around the MCT crossover. It was also found that the corresponding static length scales grow mildly and monotonically across the MCT, while the dynamic length scale grows rapidly[53]. MCT corresponds to a spinodal point of the loss of the metastability of the RFOT mosaic tiles. Therefore, it is likely that there exist more than one type of structural relaxation mechanisms around MCT, *i.e.*, the feedback mechanism of MCT and the cooperative rearranging regimes (CRR) of Adam-Gibbs theory. However, very deep in the energy landscape, the CRR mechanism should dominate. Therefore, two length scales should presumably converge again. It should be noted that MCT mechanism normally only involves structural factor based on two-body correlation functions. Another possible explanation of the discrepancy between the static and dynamic correlation lengths might simply due to the incorrect ways they are defined.

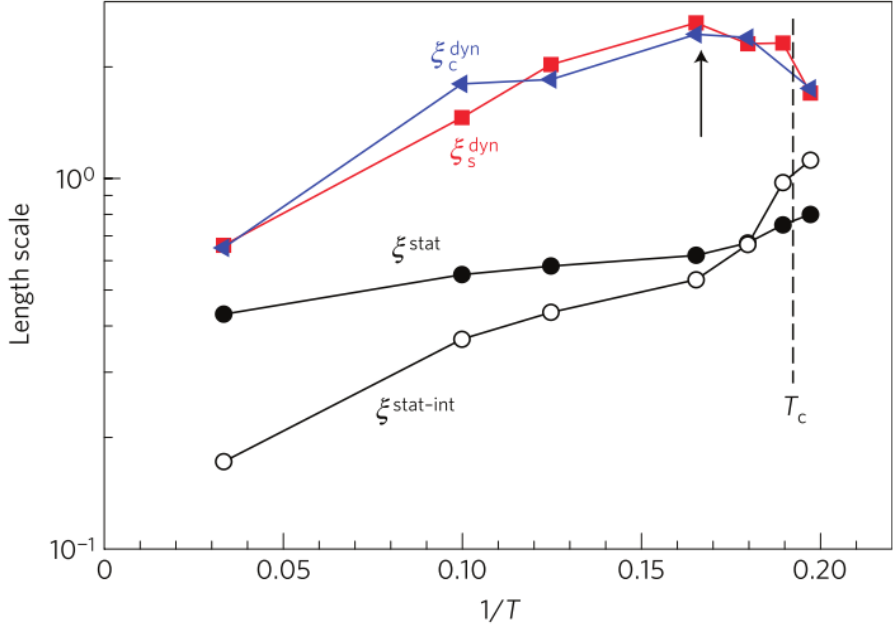

**Fig.2.** Temperature dependence of static and dynamic length scales identified. Dynamic length scales exhibit a non-monotonic behavior with a maximum at T=6.0, whereas static length scales increase modestly above $T_c$. The figure is reproduced from[53].

It was recently suggested that the structural relaxation or α-relaxation time is mainly determined by the growth of the static length scale rather than the dynamic length scale based on finite-size scaling[54]. However, it should be cautioned that most of the conclusions are drawn based on simulations on model glass formers over a very narrow temperature range where the static length scale is not significantly different from the average interparticle distance.

### 3. Granular system as a hard-sphere model glass former

Hard-sphere glass system is a popular model glass former since it can simulate systems like granular systems, colloidal suspensions and metallic glasses[55]. The dynamic arrest in glassy systems is generally believed to be mainly caused by the short-ranged repulsive interactions induced caging effect. However, it should also be noted that attractive interactions can have significant influence on the dynamics and therefore the glassy physics [56, 57]. Therefore, in addition to dry granular systems, we will also cover studies on wet granular packings to simulate glassy systems with attractive interactions.

### 3.1 Properties of granular materials

Experimental wise, granular system remains one important model hard-sphere glass former as colloidal suspensions[58]. Nevertheless, it also has many unique properties which differentiate it from an ideal one. Due to the irrelevance of the thermal energy and inelastic collisions between particles, granular systems are ordinarily at zero temperature[6]. Additionally, there exists friction between particles. Without external energy input, the system is ordinarily in an out-of-equilibrium amorphous solid state. To achieve an out-of-equilibrium steady state or access the glassy dynamics, energy has to be fed into a granular packing in order for it to explore the phase space. The system can be agitated by continuous vibrations, discrete taps or bulk forces like gravity such as in a gravity-driven flow[59-61]. For 3D granular systems, it has been observed that the system under discrete taps can evolve to a thermodynamic steady state when the final packing fraction is solely dependent on the tapping intensity instead of the preparation history[61]. Also, it is also popular to prepare a "thermal" system using 2D vibrated or air-blown granular systems since gravity does not play a role[62, 63]. Granular particles are much larger than colloidal particles, the control of size dispersion and tracking of the particle coordinates can be easier than those in colloidal experiments[64].

Since granular packing is normally at zero temperature, more interest has been focused on jamming related properties like force chains[65], soft modes[66], scaling behaviors of jamming transition[17]. Since our interest of current work is on glass transition and especially on glass structures instead of jamming transition, we will not go into details on these topics.

### 3.2 Glass and jamming transitions

Not unique for granular systems, the confusion over the relationship between glass and jamming transitions has puzzled the research of hard-sphere glass systems for quite some time. The jamming transition is a zero-temperature geometric transition and demonstrates interesting properties like marginal stability and critical scaling behaviors[17]. Whether the underlying structure plays a role in jamming transition remains unknown. Before, the relationship between jamming and glass transitions was not clearly understood, and jamming mechanism was also considered as one candidate theory for glass transition. In previous studies on granular systems, there exists a mixed use of both concepts[62].

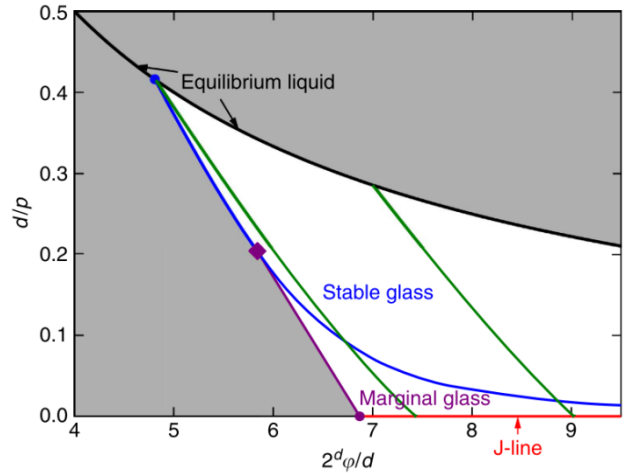

**Fig.3.** Phase diagram of infinite-dimensional amorphous hard spheres. Pressure p-packing fraction $\varphi$ phase diagram for $d \rightarrow \infty$ hard spheres. The white region indicates the regime where the (meta)basin structure is present, either as a simple stable glass or as a marginal fractal glass. The left-most boundary of the glass region is the threshold line. The 'J-line' of jammed packings is found along $p = \infty$, which always falls within the marginal phase. Although solving the mean-field out-of-equilibrium dynamics of hard spheres remains an open problem, an adiabatically slow compression should leave the equilibrium liquid line and eventually reach the J-line, while remaining within the white region. The green lines are two examples of an adiabatic following of a glass state. The figure is reproduced from[67].

Subsequently, distinction between jamming and glass transitions has been clarified from both thermodynamic[68] and rheological[69] points of view. A recent theoretical approach has tried to incorporate the jamming transition into the framework of glass transition through a Gardner transition to a fractal free-energy landscape[67]. Therefore, it is expected that this could bring a unified theoretical framework to both transitions. Specifically, the hard-sphere system undergoes glass and Gardner transition as it probes deeper and deeper in the landscape. The Gardner transition corresponds to sampling the fractal subbasins within the glass metabasin. It is therefore natural to expect the high-temperature glass transition vestiges should be encoded even in zero-temperature glassy systems like static granular packings, especially, its structural signatures. However, it should also be noted that although zero-temperature granular systems can inherit high-temperature glass structures, once perturbed weakly, it will behave very differently from a finite-temperature thermal glass. This is both due to the complex role played by friction and the fact weak perturbation will likely only move the system around the bottom of the fractal landscape basin. Therefore, significant different dynamics or rheological behavior is expected from thermal glassy systems. One other interesting fact drawn from above study[67] is that, close to the MCT transition, both the glass and Gardner transitions happen at almost the same reduced temperature, therefore, it would be likely to attribute some hallmark glassy behaviors previously observed at MCT to the Gardner transition and jamming phenomena, which adds another twist to the relationship between two transitions.

### 3.3 Glass transition in 2D granular systems
### 3.3.1 Experimental techniques

The investigation of glassy dynamics using quasi-2D monolayers of sheared, shaken, and fluidized grains is important[52]. Durian *et al.* developed a gas-fluidized granular setup that the shedding of turbulent vortices makes the system thermal-like[70-72]. Quasi-2D granular systems that vibrated vertically[63] or horizontally[73] are widely investigated. To have a better thermal bath, Olafsen *et al.* introduced a dimer layer between the monomers and the vibrated plane so that the monomers are driven by collisions with dimers randomly while the dynamics of dimers is spatially correlated[74]. Dauchot *et al.* also found that bidisperse granular disk packing behaves the same way as glassy system under quasistatic cyclic shear[75].

### 3.3.2 Glass transition in 2D granular systems

Keys *et al.* investigated the dynamics in a system of air-driven granular beads[62]. As density of the system increases, the structural relaxation slows down and dynamics becomes heterogeneous similar to atomic and molecular glassy systems. As the system approaches J point, both characteristic time and dynamical correlation length diverge. The structure of dynamic heterogeneity has also been analyzed to have similar "strings" as previous simulation on model glass formers[76]. Dauchot and collaborators carried out cyclic shear experiment in a monolayer of bidisperse disks. They found similar caging dynamics and dynamic heterogeneity as those in ordinary glass formers[77, 78]. It is proposed that the dynamical heterogeneity originates from the facilitation and avalanche build-up of clustered cage jumps[79].

To provide a structural mechanism for dynamic heterogeneity, Tanaka *et al.* observed medium-range crystalline order in a vibrated 2D granular system[63]. The spatial distribution of this order is closely correlated with the clusters of slow particles. By comparing the correlation length of the order and dynamical correlation length, they found that the values of two lengths are very close and tendencies are almost the same, indicating that medium-range crystalline order is also related to the dynamical heterogeneity.

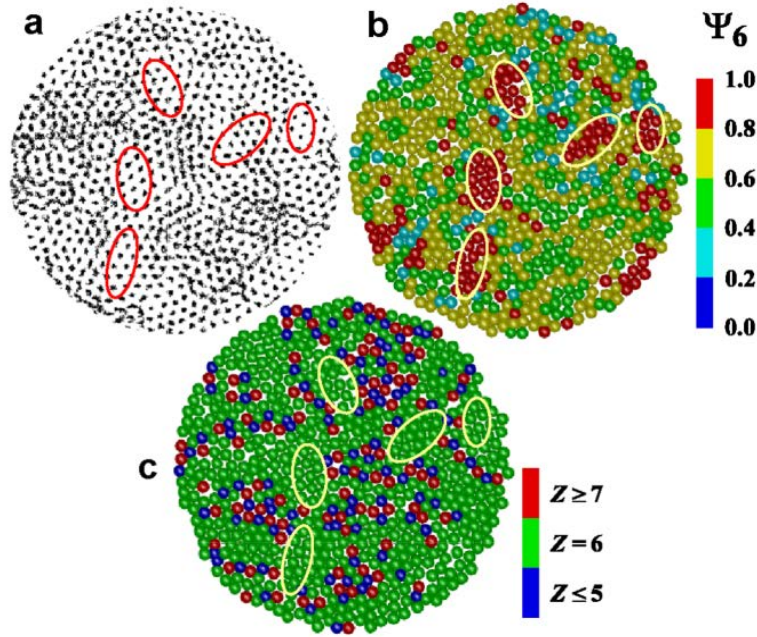

**Fig.4.** Relationship between dynamical heterogeneity, medium-range crystalline order, and topological defects for $\varphi = 0.774$. (a) Particle trajectory. (b) Spatial distribution of the time-averaged bond-orientational order parameter. (c) Spatial distribution of the coordination number Z. The figure is reproduced from[63].

## 3.4 Glass transition in 3D granular systems
### 3.4.1 Experimental techniques

Due to the opaqueness or strong scattering properties of granular particles, direct optical visualization of 3D granular structures is impossible. In addition to refractive index matched imaging[80] and magnetic resonance imaging (MRI)[81], X-ray computed tomography (CT)[82-86] can be used to measure the internal structure of granular packings. After the rapid development in the last several decades, nowadays CT technique transforms from a qualitative diagnostic tool to a quantitative one. CT is non-invasive and non-destructive. This is to be compared with refractive index matched method where the particles have to be submerged in specialized fluid. MRI normally has lower spatial resolution (~10 microns range)[87] and costs more than CT. CT not only can be applied for the study of granular packing, but also colloidal and foam packing[88]. Complex packing that contains multiple components (*e.g.* wet packing) is also available since CT is based on the absorption dependency of X-rays on material density and the reconstructed images contain the density spatial distribution information [89, 90].

The spatial resolution of CT is improved from millimeter to micrometer (μCT)[82, 91], or even to nanometer in synchrotron radiation recently[92-94] so that a wide size range of particle from the micrometer to centimeter is accessible. Due to the rapid increase of the power of the X-ray tube or the photon flux in synchrotron radiation, the scan time of CT is reduced from hours to minutes (commercial μCT), or to seconds (medical CT or μCT in synchrotron radiation[91, 95]). When the X-ray flux from the white-beam instead of the monochromatic beam is utilized in synchrotron radiation[96], the time resolution of subsecond can be achieved by 4D CT technique[97, 98] to track granular system's evolution under perturbation.

To access individual particle centroid and size, image segmentation techniques have to be applied. The tomography-reconstructed 3D images are normally analyzed by a marker-based watershed imaging segmentation technique[99-101]. The watershed method can determine the centroid with a precision of around 3% the voxel size. As a comparison, index matched method can reach 0.1 voxel size[102]. With a detector pixel size around 6.5μm, the centroid of a 200μm particle can be determined within 0.2μm. Similar spatial precision is obtained based on intensity correlation method with a reference sphere[103].

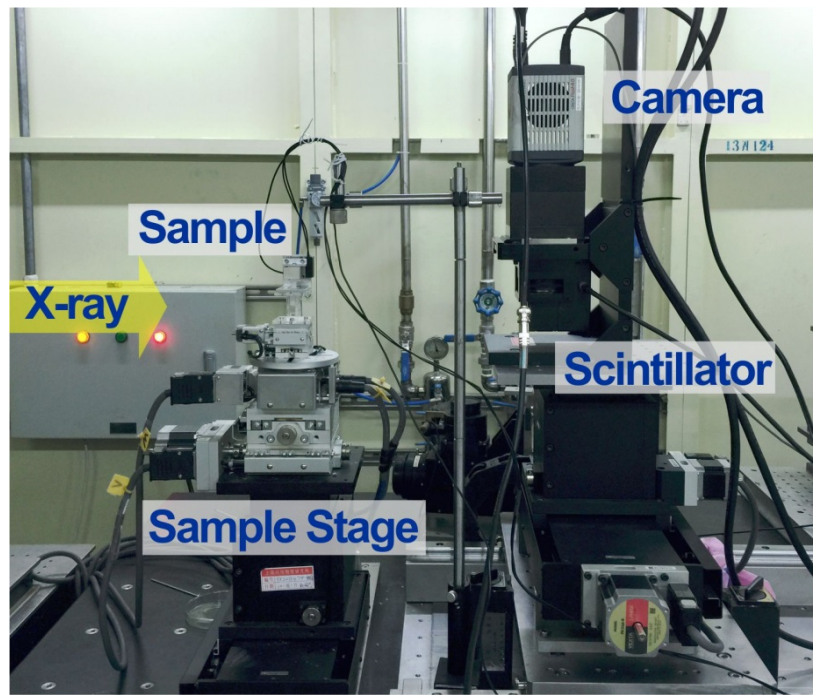

**Fig.5.** The synchrotron tomography set-up of granular packing at BL13W1 beamline of Shanghai Synchrotron Radiation Facility (SSRF).

**3.4.2 Glass transition in dry granular packings**

In 3D, the glassy dynamics is less studied due to the experimental difficulty. Most of exisitng experimental works have been carried out using X-ray tomography techniques to study the static packing structures. Aste *et al.* provided the first systematic measurement of the disordered 3D sphere packing structures, including geometric quantities like pair correlation functions, contact number, bond orientational order and local free volume [103]. Combined with entropy analysis, they found that the increase of quasi-tetrahedra is accompanied by the decrease of the entropy, and the entropy putatively will reach zero around packing fraction 0.66 on extrapolation[42]. Later on, Francois *et al.* studied the crystallization of sphere granular particles[104]. They found that tetrahedral structures of particles whose population saturates at random close packing density, and decreases with the emergence of crystalline phase since they are geometrically frustrated.

Recently, using granular packing as a model hard-sphere glass, Xia *et al.* carried out a systematic study involving structures, thermodynamics and dynamics[41]. In this work, a "hidden" polytetrahedral order has been identified as the glass structural order parameter. The order is found to be spatially correlated with the slow dynamics. As shown in Fig. 6, it is geometrically frustrated and has a peculiar fractal dimension. Additionally, as the packing fraction increases, its growth follows an entropy-driven nucleation process similar to that of the RFOT theory[16, 105]. In this work, the complexity or configurational entropy is calculated based on the Edwards' volume ensemble, which recently has been proven to be exactly valid for granular packings after many years of debate over it applicability. This work essentially elucidates the structural nature of the RFOT mosaic tiles, thus relating frustration limit domain theory with RFOT theory. Interestingly, the surface dimension $\theta_s$ of polytetrahedra measured according to the mosaics model is consistent with the fractal surface dimension of polytetrahedral clusters, whose value lies in the range of 2.2-2.5[106] instead of $\theta_s = 2$ or $\theta_s = 1.5$ by original RFOT theory[107].

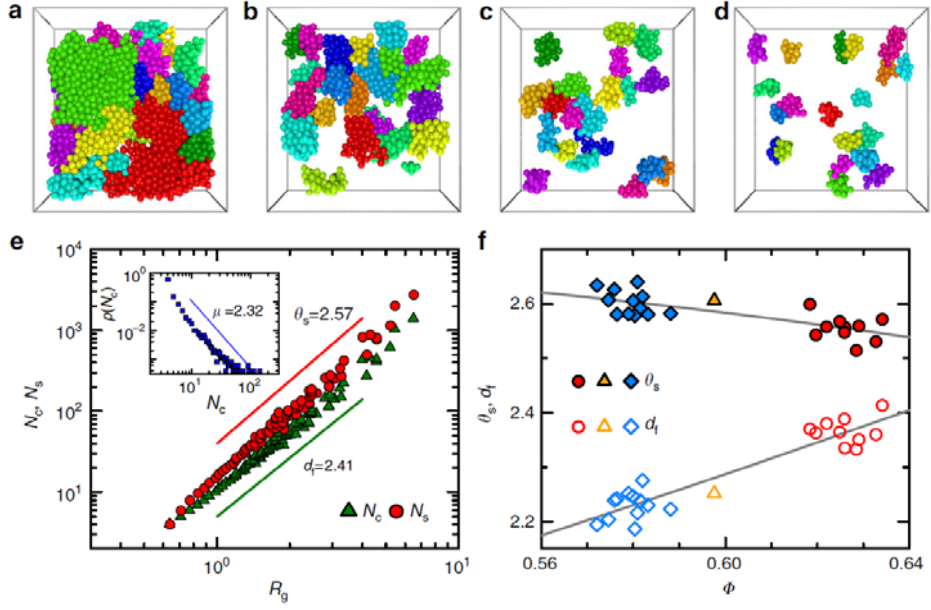

**Fig. 6.** Structure of the polytetrahedral order. (a-d) Configurations of the largest twenty polytetrahedral clusters for packing of (a-d) $\Phi$ =0.634, 0.618, 0.598 and 0.572. (e) The surface area $N_s$ (circles) and cluster sizes $N_c$ (circles) versus the radius of gyration $R_g$, $N_s$ is defined as the number of tetrahedra face-adjacent to a polytetrahedral cluster formed by quasi-regular tetrahedra. The inset shows the probability of finding a cluster with size $N_c$. Only data for $\Phi$ =0.634 is shown. The solid lines mark the slopes of the corresponding scaling behaviors: $N_s \propto R_g^{\theta_s}$, $N_c \propto R_g^{d_f}$ and $p(N_c) \propto N_c^{-\mu}$. (f) $\theta_s$ and $d_f$ versus $\Phi$. The figure is reproduced from[41].

Other structural motifs with five-fold symmetry such as full icosahedron are also suggested to play important roles in glass transition especially in metallic glass[32, 43, 108]. Xia *et al.* carried out a systematic investigation of potential role played by a family of local structural motifs with five-fold symmetry[109], *i.e.*, icosahedron, pentagonal bipyramid (7A), bipyramid (5A), and tetrahedron. As shown in Fig. 7, the spatial correlation of family of structural motifs with five-fold symmetry can be analyzed through a similar percolation analysis of quasi-regular tetrahedra[41]. However, nearby structural motifs may share different numbers of particles, and strong spatial correlations beyond random percolation are only observed when these local structural motifs are connected in very specific interlocking ways, *i.e.*, neighboring icosahedra that share a seven-particle pentagonal bipyramid, shows the strongest spatial correlation compared to icosahedra sharing other different numbers of particles. Similar correlations were found for interlocking pentagonal bipyramid and bipyramid structures, and the face-sharing tetrahedra in the previous work just corresponds to two tetrahedra sharing three particles. This essentially illustrates a hierarchical connection scenario, in which the correlation between large structural motifs with five-fold symmetry can essentially be decomposed down to spatial correlation based on face-adjacent quasi-regular tetrahedra.

The fact that the decomposition scheme works suggests that correlation calculations based on different structural motifs are very similar in nature. However, there exist subtle differences which can lead to different interpretations of the growth mechanism of glass order. If we assume that as packing fraction increases, the structural order grows, the icosahedron-like order will percolate from an almost empty interstitial "liquids", while the tetrahedral order grows through a gradual merging of small mosaics since almost all particles are already involved in at least one quasi-regular tetrahedron even at the lowest packing fraction. These are two rather distinctive candidates for the structural mechanism of glass transition. However, since what reside in the "liquids" of quasi-regular icosahedra are mostly quasi-regular tetrahedral structures instead of randomly packed liquid-like structures, therefore the mosaic-merging picture is more likely.

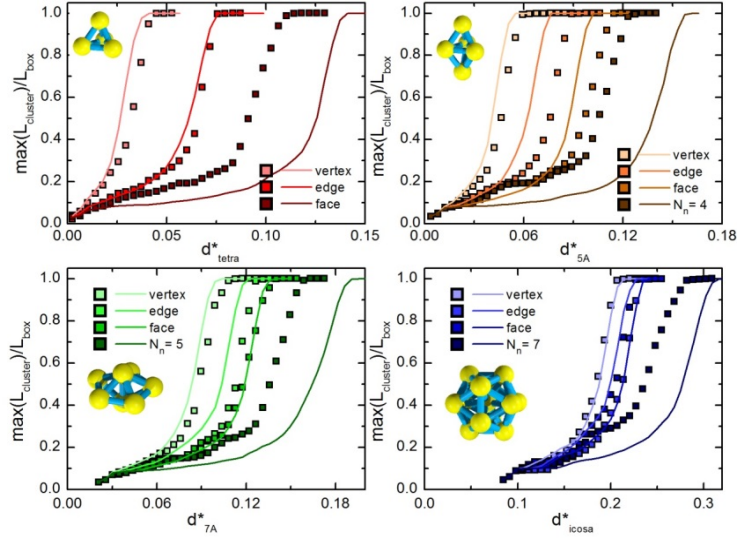

**Fig.7.** The order percolations (squares) and random percolations for (a) tetrahedral, (b) 5A, (c) 7A and (d) icosahedral orders for packing with $\Phi = 0.634$. $L_{box}$ is the size of the cubic region cut out of the packing. $L_{cluster}$ is the longest spanning range of each cluster in directions parallel to the three axes of the cube, and $\max(L_{cluster})$ is the longest among them. Smaller percolation threshold of order percolation than the corresponding random one suggests spatial correlation of the structural motif under some specific connection criterion. In each plot, data points with increasing darker colors represent increasing number of particles shared by neighboring structural motifs $N_n$. $N_n = 1, 2, 3$ correspond to vertex-, edge-, and face-sharing connections respectively. For $N_n > 3$, only percolations with evident differences with their random counterparts are shown, which is $N_n = 4$ for 5A, $N_n = 5$ for 7A and $N_n = 7$ for icosahedral orders. The figure is reproduced from[109]

Cao *et al.* studied the friction-induced bridge structures in granular packing[99], which are arch-like structures observed in flowing granular materials which are supposed to be force-bearing. As noted later[41], the bridge structures bear great resemblance to the polytetrahedral glass orders. The presence of friction might help to stabilize these structures in real granular packings and suggests their proximity to mechanical rigidity.

### 3.4.3 Wet granular packing as attractive glass

For hard spheres with short-range attractive interaction, MCT predicts a re-entrant glass transition from repulsive glass to attractive glass [110]. Simulations[111] and experiments on colloidal systems[112-114] supported this prediction. These studies distinguished repulsive and attractive glasses mainly based on dynamical and mechanical properties instead of structure. By adding a small amount of liquid into granular packing, the liquid will form capillary bridges which generate attractive interactions between particles. One can then use wet granular packing to study attractive glass. Wet granular packings show drastically different mechanical behaviors from dry ones[90].

Trying to understand the structural difference between an attractive and a repulsive glass, Li *et al.* compared the structures of wet and dry granular packings with similar packing fraction[39]. The microscopic structures of wet granular packings look very similar to those of colloidal gels[38]. There exist many locally favored structures (LFS) which are parts of an icosahedron, such as bipyramids, pentagonal bipyramids, *etc*. The fractions and types of LFSs in wet granular packings are similar to those in colloidal gels suggests that the analogy is not superficial and is related to a possible shared dynamic arrest mechanism of structural origin. The initial porous gel structure corresponds to a "liquid–gas" coexistence[115]. The large pores correspond to the "gas" phase and the dense particle network is similar to the "liquid". The compaction facilitates the phase separation by removing the "gas" phase until the system arrested into an attractive glass around a packing fraction of 0.6. The number of spheres in LFSs changes relatively slightly upon compaction which supports the phase separation scenario. Very regular LFSs on the other hand can seldom be found in dry packings with similar packing fractions[39, 41]. This work therefore yields a direct evidence of the structural differences of an attractive glass from a repulsive one, where both caging and bonding effects are at work. A possible explanation of the structural difference is that the large bonding force will frustrate growth to large RFOT mosaics since it will always outweigh the free energy gain from entropic effect. Correspondingly, the LFSs in dry packings cannot maintain very regular shapes because fluctuations are large, but they can grow to larger RFOT mosaics. However, both scenarios support a structural mechanism of dynamic arrest by LFSs with five-fold symmetries.

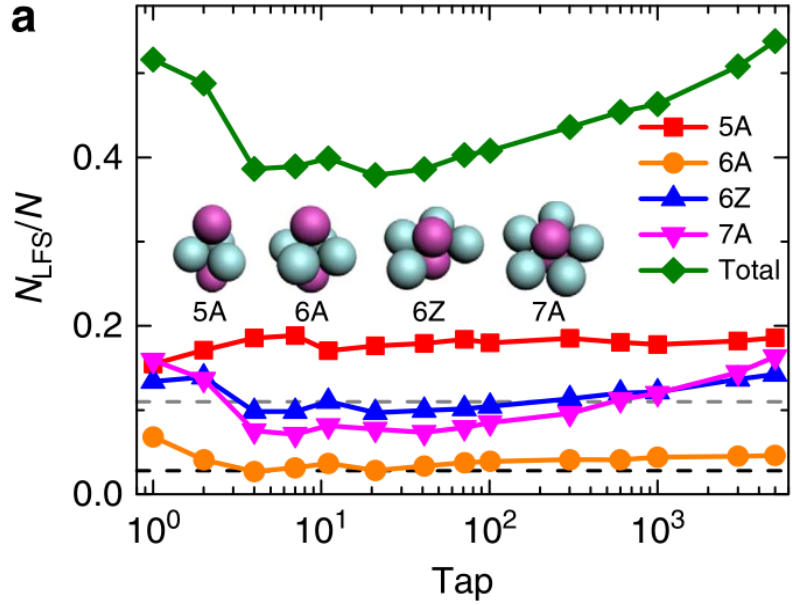

**Fig. 8.** The fractions of LFSs inside the wet packing as a function of tapping number. The figure is reproduced from[39].

### 3.4.4 Nonspherical granular packing

Granular matter in nature is composed of particles of different sizes and shapes. Among particles of all different shapes other than spherical ones, ellipsoid particle is one of the simplest. Particularly, elongated ellipsoids with aspect ratio around 1.4 have a random close packing density around 0.734, much larger than the RCP limit of sphere packing and is close to that of FCC close packing[116], which suggests a possible ideal glass state of non-spherical particles. Similar large packing fraction beyond RCP has also been observed in polydisperse particle packings.

The local packing structure of 3D ellipsoid packing was obtained by Xia *et al.* using medical X-ray tomography[117]. It was clearly demonstrated that the particle asphericity induces an effective polydispersity effect to influence the packing properties. The local structure of ellipsoid packing was explicitly mapped onto a polydisperse spherical one, and a "granocentric" model was hereafter applied to the effectively polydisperse packing and reproduced most of the experimental observations[118]. The "granocentric" approach is essentially a mean-field model which neglects correlation beyond first shell, and the relative position of neighboring particles. This essential maximization of the entropy associated with angular relative positions between particles suggests weak structural correlation of the packing. As hard-sphere system goes deep in the free energy landscape, it will develop angular correlations which results from an effective attractive interaction. The absence of correlation suggests that nonspherical particles might not be able to develop orders as easily as spherical ones.

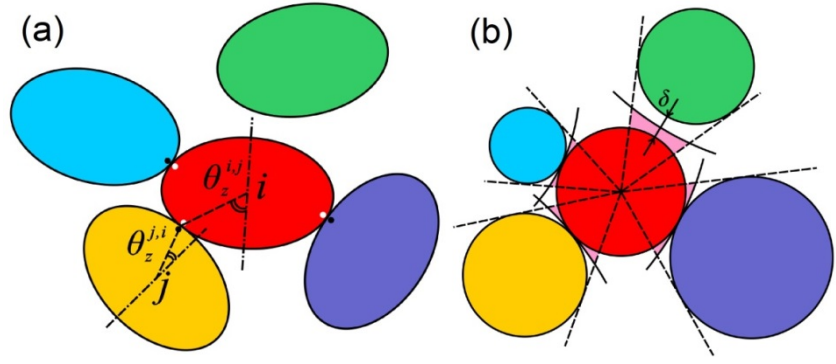

**Fig. 9.** Schematic diagrams of a local contact configuration and the mapping to a polydisperse sphere packing model. The figure is reproduced from[117].

## 4. Conclusion

Granular systems are very useful model systems to study glass transitions. In the current review, we have provided strong evidence that monodisperse granular packings with purely repulsive or attractive interactions, the glass order is tetrahedron or other structural motifs with five-fold symmetries. The growth of the order follows specific geometric rules which have been overlooked in previous studies. Additionally, the RFOT mosaics have fractal shapes but form an overall compact structure with fractal domain walls among them. This is different from previous speculation that the glass order will form a rather open fractal structure. Nonspherical or polydisperse particle packings behave more mean-field like which suggests there exist insignificant correlations in these systems. Future work will be devoted to understanding how glass order influences the mechanical properties of glass materials, *e.g.*, plastic deformation, avalanche, shear band, *etc*. It is also expected that a deep understanding of model granular systems will also greatly benefit our understanding on related topics like granular rheology. One last thing to emphasize is that the study of 3D structure and dynamics in granular materials is crucial and new experimental tools have to be developed.

**Acknowledgments**

The work is supported by the National Natural Science Foundation of China (No. 11175121, 11675110 and U1432111), Specialized Research Fund for the Doctoral Program of Higher Education of China (Grant No. 20110073120073).